\documentclass[12pt,preprint2]{aastex}

\newcommand {\tighttable}{\def\baselinestretch{1.0}}

\newcommand {\cc}{$\rm cm^{-3}$}
\newcommand {\cl}{$\rm cm^{-2}$}
\newcommand {\ltsim}{\raisebox{-.5 ex}{$\;\stackrel{<}{\sim}\;$}}
\newcommand {\gtsim}{\raisebox{-.5 ex}{$\;\stackrel{>}{\sim}\;$}}
\newcommand {\kms}{${\rm km\,s}^{-1}$}
\newcommand {\s}{${\rm s}^{-1}$}

\newcommand {\lya}{Ly~$\alpha$}

\newcommand {\h}{\ion{H}{1}}

\def\arcsecpoint{\ifmmode ''\!. \else $''\!.$\fi}
\newcommand {\he}{\ion{He}{2}}

\newcommand {\frache}{$x_{\rm HeII}$}
\newcommand {\frach}{$x_{\rm HI}$}
\newcommand {\ciii}{\ion{C}{3}}
\newcommand {\civ}{\ion{C}{4}}
\newcommand {\lm}{$\lambda$}

\newcommand {\qa}{HE2347$-$4342}
\newcommand {\qb}{HS1700+6416} 
\newcommand {\qc}{Q0302$-$003} 
\newcommand {\qd}{Q2311$-$1417}
\slugcomment{To appear in {\it The Astrophysical Journal}}
\shortauthors{Zheng} 
\begin{document}
\title{Development of Hydrogen and Helium Proximity Zones around Quasars} 
\author{
Wei Zheng\altaffilmark{1},
Avery Meiksin\altaffilmark{2},
and 
David Syphers\altaffilmark{3}
} 
\altaffiltext{1}{Department of Physics and Astronomy, Johns Hopkins 
University, 3701 San Martin Dr., Baltimore, MD 21218, USA} 
\altaffiltext{2}{Scottish Universities Physics Alliance (SUPA); Institute for Astronomy,
 University of Edinburgh, Royal Observatory, Edinburgh EH9 3HJ, United Kingdom}
\altaffiltext{3}{Physics Department, Eastern Washington University, Science 158, Cheney, WA 99004, USA} 
\begin{abstract}
Increasing evidence suggests that \he\ proximity profiles in the quasar spectra at $z\sim 3 - 4$  
are sensitive probes of quasar ages. But the development of their \h\ counterparts is difficult to 
trace and remains poorly constrained. 
We compare the UV spectra of 15 \he\ quasars with their high-resolution optical counterparts
and
find a significant correlation between the sizes of \he\ and \h\ proximity zones. The luminous 
quasar \qa\ displays a null proximity zone in both \he\ and \h, suggesting that it is extremely young 
(age $< 0.2$ Myr).
Three other quasars also display small proximity zones for \he\ and \h. 
There is no evidence that a \h\ ionization zone expands considerably 
faster than its \he\ counterpart. The results suggest that the expansion of quasar ionizing fronts may be 
noticeably slower than the speed of light, and raise the possibility of distinguishing 
young and old quasars from the sizes of their \h\ proximity zones.
\end{abstract}

\section{INTRODUCTION}\label{sec_intr}

The reionization of intergalactic helium took place at redshift $ z\sim 3 - 4$
as the result of increasing quasar activities. The powerful radiation of 
quasars creates high-ionization zones in their vicinity, and these 
``cosmic bubbles'' can be traced with proximity profiles of absorption 
\citep{zheng95,zheng15,khrykin} at radial distances of up to $\sim 20$ Mpc (proper distance, 
for all the distances throughout the paper). 
The large extent of proximity zones suggests that this expansion takes a long time, on the order of 
million years, as a huge amount of high-energy photons are needed for ionizing helium over 
a vast space. Gradually, the expansion of ionizing front slows down at large distances as
the quasar flux is geometrically diluted and photon-limited.

After a quasar is turned on, the expansion of its ionization zones takes place in both \h\ and \he, and the 
former is believed to propagate faster as intergalactic hydrogen has been largely ionized at $z\sim 3-4$. 
Nevertheless it would take a significant amount of time, on the order of million years, for 
the light signal to travel over a distance of many Mpc. 
The term of quasar age as discussed in this paper means the part of a quasar's 
lifetime during which it provides the ionizing photons for the proximity zones.
While it looks plausible that the sizes of proximity zones bear the signature of quasar 
ages, the possibility becomes complicated as 
the observed proximity effect along a line of sight may be considerably larger than the 
intrinsic one for a given quasar age. For hydrogen, it is 
believed that the propagation front expands near the speed of light, as the surrounding 
hydrogen has been highly ionized. As viewed along a line of sight, the 
expansion rate is magnified \citep{white}.
Some models \citep{bolton07,lu,khrykin} adopt an infinite speed, thus ruling out the 
possibility of using \h\ zone sizes to scale quasar ages. 
If that is the case, young quasars at $z\sim 3$ that display small or null proximity zones in \he\ 
would show a full-sized \h\ proximity effect. In this paper we attempt to address
whether the \h\ proximity zones at $z\sim 3 - 4$ carry information on quasar ages. 

Given the flux of metagalactic UV background (UVB) radiation field 
$\Gamma_{\rm HI} \sim 10^{-12}\ {\rm s}^{-1}$ \citep{bolton} and a 
nominal quasar luminosity above 1 Ry of $\dot N_\gamma \sim 10^{57.5}\ {\rm s}^{-1}$, the size of a \h\ 
proximity zone is $\sim 10$ Mpc at $z\sim 3 - 4$.  
For a luminous quasar such as \qb, its zone size would be over 20 Mpc.  For \he, the estimated proximity-zone sizes are
sensitive to redshifts and less certain, but probably larger than their \h\ counterparts as the \he-ionizing UVB was rapidly developing.
Indeed, the measurements of \he\ proximity zones cover a significant range. 
Increasing evidence \citep{shull, syphers13,zheng15,khrykin19} suggests that some quasars show quite small \he\ proximity zones,
implying that they are likely young quasars that have not yet built significant 
high-ionization zones in their vicinity. If the \h\ front travels at the speed of light, 
a \h\ proximity zone of a young quasar along the line of sight should display a full size of $\sim 10-20$ Mpc, 
according to current models.

\section{DATA}\label{sec_data}

Our sample consists of 15 quasars whose 
\he\ (UV) and \h\ (optical, high-resolution) spectra are available. 
The redshift ranges are between $z\simeq 2.73 - 2.95$ and $3.05 - 3.85 $.
The quasars at $2.95< z < 3.05$ are excluded because of strong geocoronal \lya\
contamination on the redshifted \he\ \lya\ in the UV data. The data of a few 
other quasars are not used 
because of too low signal-to-noise (S/N) ratios. 
All the archival UV spectra were obtained with the HST/COS instrument, except 
for PKS1935$-$692 (GHRS).
Spectra taken in multiple exposures are rebinned and then combined.  
The bin size is 0.05 \AA\ for COS/G130M grating, 0.32 \AA\ for COS/G140L and 0.14 \AA\ for GHRS/G140L. 
The \he\ proximity-zone sizes of 13 quasars in the sample were 
reported in \cite{zheng15}, and that of PC0058+0215 in \cite{worseck16} and Q2311$-$1417 in \cite{khrykin19}.
We remeasure these \he\ proximity-zone sizes at the shortest wavelengths where the proximity 
profile drops to 10\% of the unattenuated flux level (except for \qb, see \S \ref{sec-ind}), 
and the results are consistent with the previously 
reported values. For \qb\ and SDSS2346$-$0016, their \he\ proximity profiles are affected by geocoronal airglow emission, therefore their spectra were 
extracted from the portion of orbital night, using standard {\tt calcos} pipeline \citep{hodge}. 

The Keck/HIRES spectra of seven quasars are retrieved from the Keck archive and the VLT/UVES 
spectra of eight quasars are retrieved from the ESO archive. These optical spectra cover a 
wavelength range in the rest frame from 1250 \AA\ down to at least 1135 \AA. All the optical spectra are obtained in 
the form of extracted spectra, except for \qa\ in which a normalized UVES spectrum \citep{zheng04} is in hand.
The archival HIRES data have a gap between $\sim 5217$ and 5247 \AA, and
the UVES data between $\sim 5158$ and 5222 \AA. These gaps  
affect a small portion of the wavelength coverage at $z \gtsim 3.3$.
Normalized Keck/HIRES spectra for several quasars are available from the 
KODIAQ database \citep{omeara17}.
Table \ref{tbl-data} lists the quasar redshifts, their data sources and exposure times. 

The optical spectra of different orders and exposures
are rebinned to a pixel scale of 0.05~\AA\ and combined.
Absorption features are identified and measured through local troughs and continua after data smoothing by 
5 pixels. Depending on data quality, the detection limits vary 
from an equivalent width (EW) of $\sim 0.005$ to 0.03 \AA\ in the restframe 
(and hereafter unless stated otherwise). The spectral regions 
free of absorption lines are fitted with high-order polynomials, and a normalized spectrum is 
 produced. Note that our EW measurements of absorption features are not sensitive to this 
normalization process.
In the UV data, if more than one dataset exists, they are rebinned 
and then combined. The data points longward of the \he\ lya\ 
wavelength are fitted with a power law and a reddening curve with a E(B-V) value \citep{ext} and a
reddening curve of $R_V=3.1$ \citep{ccm}, in order to normalize a UV spectrum.

In the optical \h\ spectra at $z \sim 3$, \lya\ forest lines are not strong enough to form 
a continuous absorption profile, and the 
proximity effect is reflected as a decline of absorption-line numbers toward the \lya\ wavelength 
\citep{bdo}. Not all absorption lines
respond sensitively to enhanced ionizing radiation as strong \lya\ lines are 
saturated. According to the curve of growth at a nominal Doppler parameter of $b=30$ \kms,
only the EW of \lya\ lines at \ltsim 0.2 \AA\ (column density
$\ltsim 10^{13.8}$ \cl) decreases inversely with an increasing ionizing flux.
Therefore our study of the \h\ proximity effect only includes the \lya\ absorption lines 
of EW $<0.2$ \AA. Metal absorption lines are identified by ratios of their doublet 
wavelengths and narrow widths ($b < 15$ \kms) and then excluded.

\section{RESULTS}\label{sec_res}

\subsection{Estimate of \h\ Proximity Zones}

\begin{figure}
\plotone{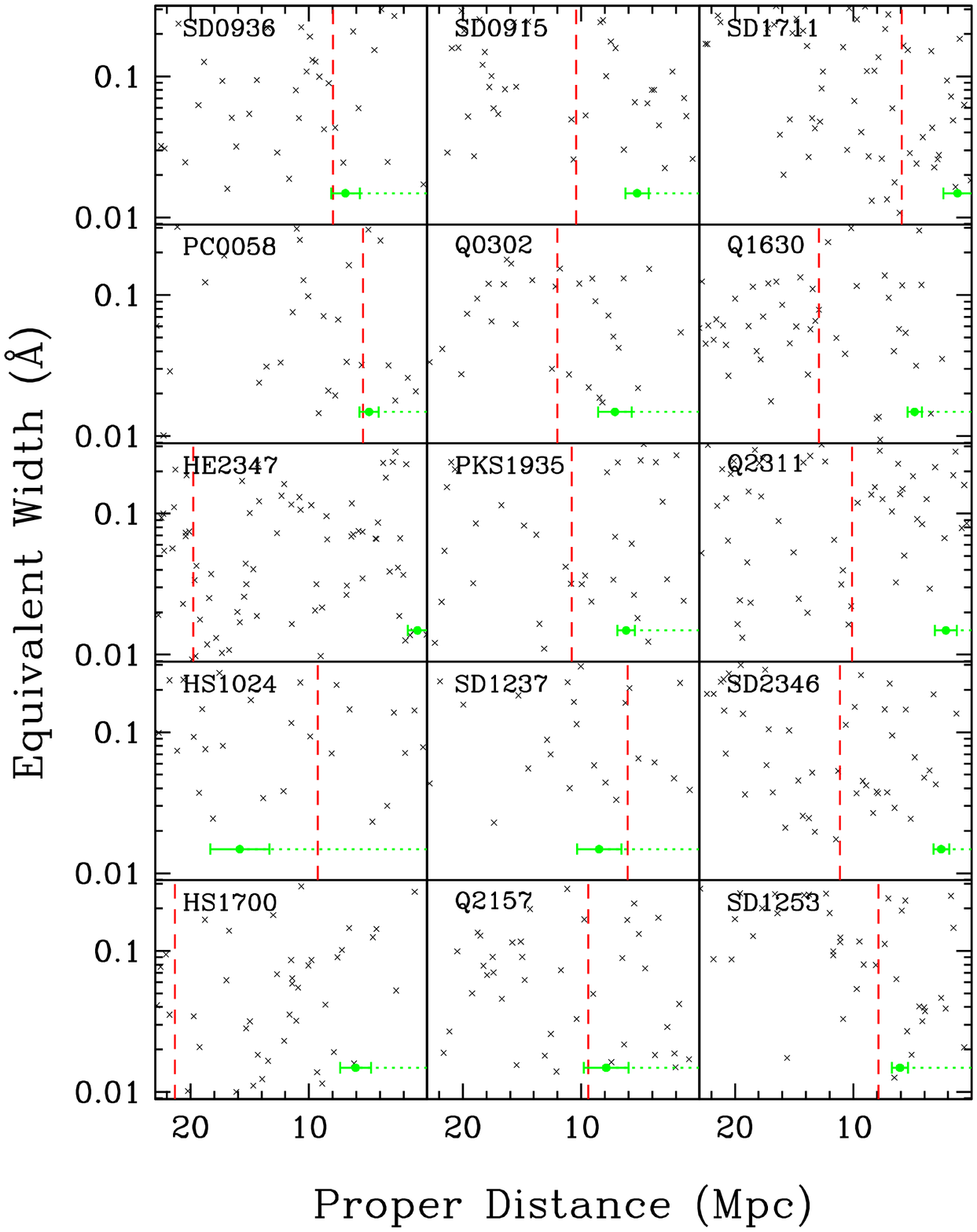}
\caption{
Distribution of \lya\ absorption lines in quasar vicinity.
The red dashed lines mark the positions for characteristic radius $R_{\omega=1}$ (Table \ref{tbl-zone}),
where the estimated quasar ionizing flux $\Gamma^{\rm Q}_{\rm HI}$ is equal to that of the UVB.
The green circles mark the end points of proximity zones with their errors. The quasars are labeled with their abbreviated names. 
\label{fig-mosaic}}
\end{figure}

In Figure \ref{fig-mosaic}  we plot the EW distribution of \h\ absorption 
lines within $ 23$ Mpc ($\approx 30$ \AA) of radial distance.
The detection limit is uneven at different parts of the 
same spectrum, as the result of line blending, instrumental sensitivity and gaps. 
Weak absorption features should be easier to find near the \lya\ peak, where \lya\ 
emission increases the flux and boosts S/N and the 
nominal \lya-line density is the highest \citep[$dn/dz \propto (1+z)^{1.28}$,][]{kim2013}.
On the contrary, the \h\ proximity effect in the quasar vicinity is seen as the
depressed number density of absorption lines and/or their strengths. 
While the proximity effect is evident in some quasars as a lack of weak absorption 
lines near the \lya\ wavelength, it is present in some other quasars as a decrease in EW 
for a similar number density  towards \lya.  For a given wavelength bin, we calculate the total EW 
as their {\it geometric} mean (the n-th root of products) 
multiplied by their number, 
to detect a decline in both the numbers 
and strengths of \lya\ absorption lines and to avoid the bias towards strong lines. 
The total EWs are calculated from wavelengths as low as 1050 \AA\ toward \lya\ at steps of 1 \AA\ 
and with bins of 3, 5  and 7 \AA. 

We compare the number density and total EW with different bin sizes to estimate the \h\ 
proximity zone in every quasar. 
If the total EW in a bin or the number of lines declines by 30\% as compared to 
those at larger distances, it is considered as a sign of proximity effect. The external effect by foreground quasars near lines of sight may complicate 
our study. The distribution of line number in the vicinity of several quasars 
is not 
monotonic, making two possible sizes of proximity zone. For example, the EW distribution in SDSS0915$-$0016 shows two dips in the distribution of
line numbers around 16 and 6 Mpc (see the upper middle panel of Figure \ref{fig-mosaic}). 
The one at a further distance is consistent with an external void, as absorption lines  
reappear at shorter distances. We therefore mark the one at 6 Mpc as the \h\ proximity zone.

\begin{figure}
\plotone{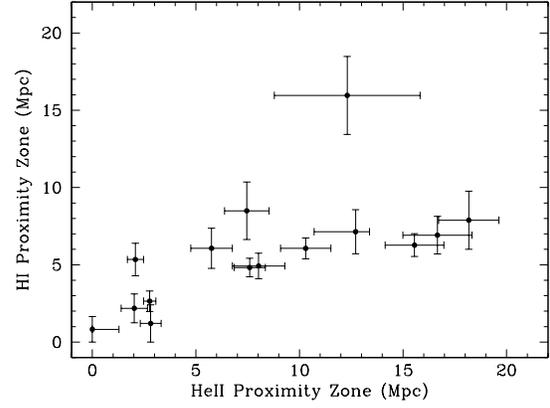}
\caption{Proximity zones of \he\ and \h\ in 15 quasars.
The errors are estimated from measurements and do not include the redshift uncertainties. 
\label{fig-heh}}
\end{figure}

\begin{figure}
\plotone{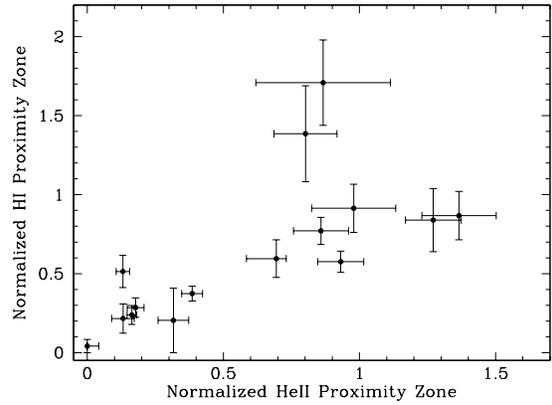}
\caption{Proximity zones of \he\ and \h, normalized by their characteristic 
sizes as derived from the quasar luminosities. 
See the caption of Figure \ref{fig-heh}.
\label{fig-norm}}
\end{figure}

The sizes of \h\ and \he\ proximity zones in these 15 quasars are listed in Table 
\ref{tbl-zone} and compared in Figure \ref{fig-heh}.
The errors are estimated from visual inspections and do not include the term 
due to redshift uncertainties, which would have the same effect on both the \h\ 
and \he\ proximity-zone sizes. In Figure \ref{fig-norm} we normalize the 
proximity zones with the expected characteristic sizes $R_{\omega=1}$,
where the estimated ionizing flux from a quasar $\Gamma^{\rm Q}$ is equal to that of the UVB.
To convert from the observed optical fluxes to the intrinsic values at 1 and 4 Ry, 
a broken power law is assumed with an index of $\alpha = -0.44$ \citep[$f_{\nu}\propto
\nu^{\alpha}$,][]{vb} below the \lya\ frequency and 
$\alpha = -1.73$ above it \citep{zheng97, telfer, lusso}. The UVB fluxes are fixed at
$\Gamma_{\rm HI} = 10^{-12}$ \citep{becker} and $\Gamma_{\rm HeII} = 
10^{-14.3}$ \s~ \citep{worseck19}. This is a rough estimate because the UV continua of quasars show significant variations,
but it serves as a reasonable mark as the proximity-zone sizes are 
proportional to the square root of UV luminosities.
Note that the proximity effect extends beyond such a characteristic distance, 
but becomes less visible.

We carry out a statistical test between the \he\ and \h\ zone sizes. It yields a Pearson 
correlation coefficient of $r= 0.62$ and a probability of no-correlation $\rho =1.2\%$. 
A test between the normalized zone sizes (Figure \ref{fig-norm}) yields similar results: 
$r = 0.69$ and $\rho =0.46\%$,  
suggest a significant correlation. 
We also carry out a linear regression between the \he\ and \h\ zone sizes. The slope of 
0.39 and a correlation coefficient 0.62 suggest that \h\ 
proximity zones are often smaller than their \he\ counterparts. 
Part of this trend is attributed to our conservative estimates of \h\ proximity zones in 
several quasars, where more than one possible size of \h\ proximity zone exist, and the larger value
is rejected because it is likely an external effect. 

\subsection{Individual Cases}\label{sec-ind}

\begin{figure}
\plotone{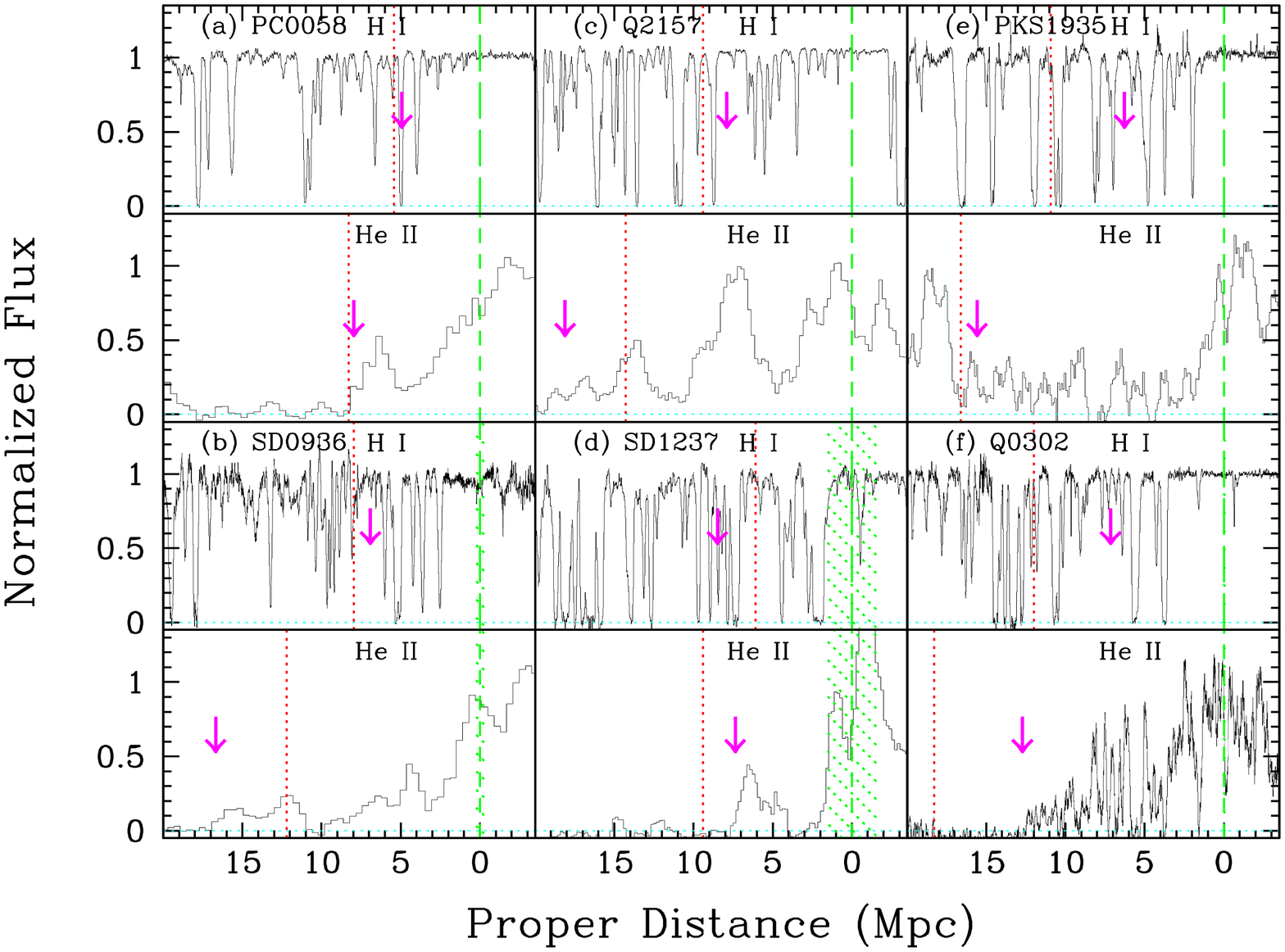}
\caption{
\h\ and \he\ spectra of six quasars, converted from wavelengths to radial distances and smoothed. 
The quasars are labeled with their abbreviated names.   
For each quasar, the \he\ spectrum is aligned and plotted below its \h\ counterpart.
The quasar positions are marked by green dashed lines, and shaded regions the range of 
redshift uncertainties. 
The magenta arrows mark the ending points of estimated proximity zones.
The red dashed lines mark the characteristic zone sizes where the quasar ionizing flux is equal
to that of the UVB. The cyan lines mark the level of zero flux.
\label{fig-mosaic3}}
\end{figure}

\begin{figure}
\plotone{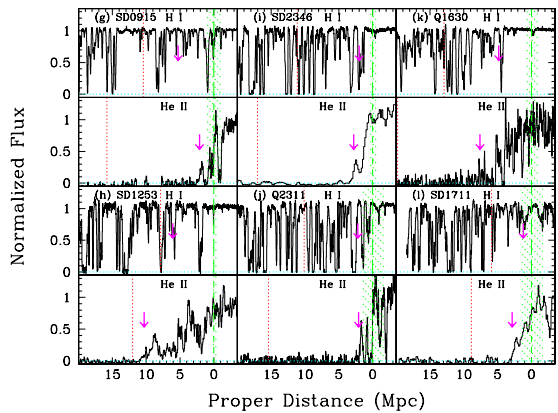}
\caption{
\h\ and \he\ spectra of six other quasars. See caption of Figure \ref{fig-mosaic3}. For
SDSSJ1711+6052, a section of the \h\ spectrum is noisy and hence not plotted.
\label{fig-mosaic4}}
\end{figure}

\begin{figure}
\plotone{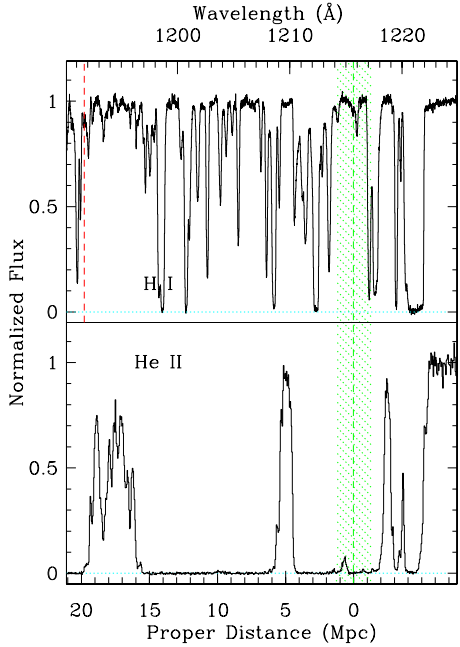}
\figcaption{G130M and UVES spectra of \qa\ ($z = 2.885$). 
The UV spectrum of \he\ is smoothed by three pixels and aligned with the optical 
counterpart. 
The quasar position is marked by a green dashed line and
a shaded region of uncertainty. Several absorption 
features are present at longer wavelengths. 
The red dashed line marks the characteristic zone size where the quasar 
ionizing flux is equal to that of the UVB.
The cyan lines mark the level of zero flux. 
There is an apparent lack of proximity effect for both \he\ and \h. 
\label{fig-2347}}
\end{figure}

\begin{figure}
\plotone{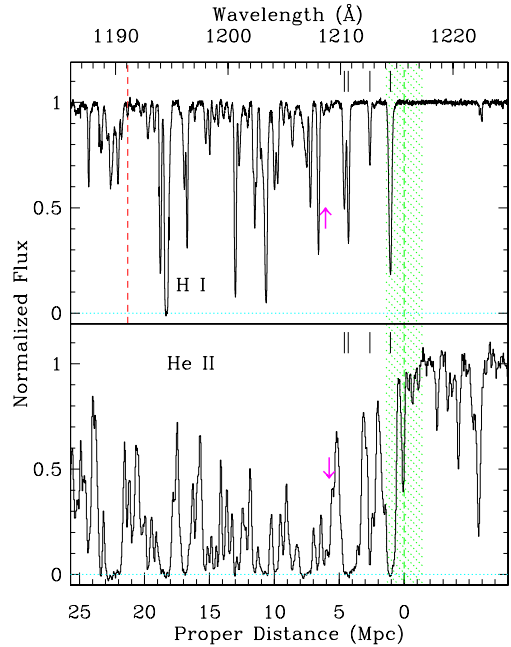}
\figcaption{G130M and HIRES spectra of \qb\ ($z = 2.748$). 
The UV spectrum of \he\ is smoothed by three pixels and aligned with the optical 
counterpart. 
The quasar position is marked by a green dashed line and a shaded 
region of uncertainty. The magenta arrows mark the respective ending points of 
estimated proximity zones.
The red dashed line marks the characteristic zone size 
where the quasar ionizing flux is equal to that of the UVB. 
The cyan lines mark the level of zero flux.
There are no weak absorption lines
in the quasar's vicinity between $\sim 1209$ and 1216 \AA, 
but four moderate absorption lines are present as marked and make the \he\ proximity 
profile choppy. Note the non-zero flux at large distances ($> 10$ Mpc). 
\label{fig-1700}}
\end{figure}

\begin{figure}
\plotone{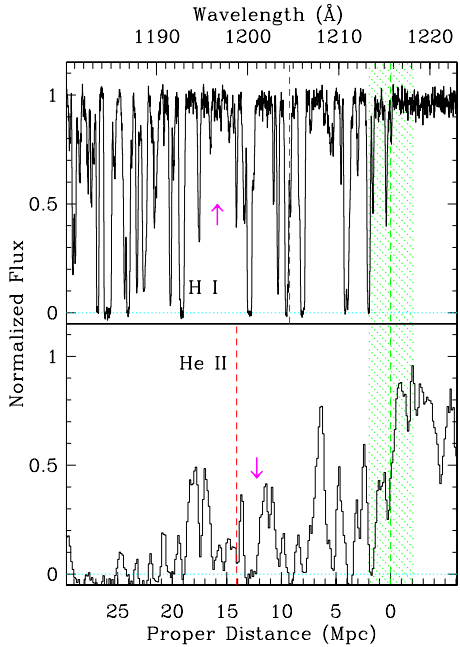}
\figcaption{G130M and UVES spectra of HS1024+1849 ($z = 2.855$).
See caption of Figure \ref{fig-2347}. 
The spectral void around 1195 \AA\ is likely of 
external cause, and a generous error in the proximity-zone estimate covers a region between 
$\sim 1198 - 1205$ \AA. \label{fig-1024}}
\end{figure}

In Figures \ref{fig-mosaic3} and \ref{fig-mosaic4} 
we plot the normalized \h\ and \he\ spectra of 12 quasars, with marks of the estimated and
characteristic proximity-zone sizes. The \he\ and \h\ spectra of three quasars with unique 
proximity zones are plotted in Figures \ref{fig-2347} - \ref{fig-1024}.

\paragraph{\qa:} 
This luminous quasar is arguably the best example for a null \he\ proximity zone 
\citep{shull}. A reliable systemic redshift of $z=2.885 \pm 0.005$ is measured from a 
broad low-ionization line of \ion{O}{1} \lm 1302 \citep{reimers}, which is consistent with a 
near-infrared spectrum \cite[R. Simcoe, private communication; see also][]{simcoe}. Several 
\he\ absorption lines are present at wavelengths longward of the \lya, up to $z_a = 2.905$, 
perhaps attributed to infalling materials. The data quality for this quasar is the highest in 
our sample as the EWs are based on a list of fitted \lya\ lines down to column density 
$\simeq 10^{12}$ \cl\ \citep{zheng04}.  
As shown in Figure \ref{fig-2347}, the density of \h\ absorption lines does not 
decrease in the quasar's vicinity, suggesting a null \h\ proximity zone ($0.8 \pm 0.8$ Mpc). 
The apparent lack of weak absorption lines around 5 \AA\ from \lya\ is attributed 
to the overlapping absorption lines between 1209 and 1214 \AA. Given a high luminosity, a fully developed proximity zone should be present beyond 
characteristic distances $R_{\omega=1} \sim 20$ Mpc for \h\ and $\sim 30$ Mpc for \he.
Actually the upper limit to the \he\ proximity zone is only 1.3 Mpc, 
less than 5\% of the anticipated value (\S \ref{sec-age}). 

\paragraph{\qb:} This luminous \he\ quasar has also been studied extensively \citep{dkz,fechner,syphers13}. The HIRES spectrum (Figure \ref{fig-1700}) reveals a proximity zone: within 7 \AA\ shortward 
of \lya, only four absorption lines at $EW > 0.1$ \AA\ and 
no weaker lines. These four lines cause deep troughs in the \he\ proximity profile, which extends to 
at least 5 \AA\ and possibly to 7 \AA\ from \lya.
Because of its high luminosity, the characteristic sizes of proximity zone are
slightly larger than that of \qa. The residual flux at $\lambda < 1205$ \AA\ 
is not zero \citep{dkz} because of the quasar's low redshift, therefore the end point of \he\ 
proximity zone is estimated to be at $\approx 5.8$ Mpc.
The measured size of \he\ proximity zone is only $\sim 17\%$ of that 
anticipated value. Overall, the proximity zone of \qb\ is clearly present, 
but significantly underdeveloped. 
 
\paragraph{\qd:} \cite{khrykin19} derived a \he\ proximity zone of $R_{\rm HeII}= 1.94 \pm 1.72$ Mpc 
(including the redshift uncertainty) for this quasar and suggested an young age of 
$< 2 $ Myr. As shown in Panel {\it j} of Figure \ref{fig-mosaic4}, absorption lines are 
present at the region shortward of \h\ \lya\ wavelength, and there is no void within 20 \AA\ toward 
the redshifted \he\ \lya\ wavelengths.

\paragraph{SDSS2346$-$0016:}
As shown in Panel {\it i} of Figure \ref{fig-mosaic4}, this quasar displays small proximity zones for \he\ and
\h\ ($\sim 2.7$ Mpc). The SDSS redshift of $z=3.489$ would place the entire proximity profile longward of
\lya, which is unlikely. We adopt  $z=3.511$ \citep{zheng15} for this quasar. Only the night portion of 
COS data is used because of airglow emission around the observer's wavelength of 1360 \AA. 

\paragraph{\qc:} 
This quasar is known for the first detection of  \he\ Gunn-Peterson trough \citep{jakobsen} as well as the first confirmation of a proximity profile \citep{hogan}. 
It represents a good example for 
its significant proximity zone \citep{heap,syphers14,khrykin}.
Its redshift is $z=3.2860 \pm 0.0005$ as derived from narrow [\ion{O}{3}] lines \citep{syphers14}. 
Optical spectra reveal a large spectral void at $z\sim 3.17$, which is attributed to the transverse 
proximity effect of a foreground quasar near the line of sight \citep{dobrzycki}.
There is an apparent lack of \h\ absorption lines within 15 \AA\ from the 
\lya\ wavelength (Panel {\it f} of Figure \ref{fig-mosaic3}), in spite of significantly higher S/N 
ratios near the \lya\ emission peak. We take the size of this \h\ void as that 
of the proximity zone. There is a void at approximately 11.4 Mpc (1200 \AA) possibly due
to another foreground
quasar Q0301$-$005 \citep[$z\sim 3.23$,][]{syphers14}. We extend the lower error of the \he\ 
proximity-zone size to reflect this possible external effect.

\paragraph{HS1024+1849:} 
This quasar has been observed with the SDSS, and the SDSS value of $z=2.8423$ seems too low as 
the most significant part of the \he\ proximity profile would have been missed. 
We fit the SDSS spectrum with a power-law continuum and Gaussians emission 
lines over the ranges without strong absorption lines. The \ciii\ redshift is a low 
value of 2.839, \civ\ 2.864 and \lya\ 2.852, using the wavelength values $\lambda_{lab}$ 
in \cite{vb}. We take their average as $z= 2.855 \pm 0.008$. 
The UV spectrum of this quasar, as shown in Figure \ref{fig-1024}, displays
prominent high-flux points within wavelengths $\sim 20 $ \AA\ from the \lya.
The optical spectrum is choppy near \lya, with several small troughs that match 
the strong absorption lines in the optical spectrum. However, the number of 
absorption lines and their total EW per wavelength bin of 5 \AA\ is steady within
$\sim 20$ \AA\ of \lya\ then increase towards shorter wavelengths. We consider 
a large \h\ proximity zone for this quasar.

\paragraph{PKS1935$-$692:}

\cite{anderson} reported a wide, flat \he\ proximity "shelve'' shortward of the \he\ 
\lya\ wavelength, extending to at least about 14 Mpc. As shown in Panel {\it e} of 
Figure \ref{fig-mosaic3}, there is a significant, real void at about 18 Mpc. Because of the contrast 
between this void and the rest of the proximity profile, we consider it an external feature.

\subsection{Redshift Uncertainties}\label{sec_dz}

A significant factor that affects the proximity-zone sizes is the redshift 
uncertainties. For the three
bright quasars listed at the top rows of Table \ref{tbl-data}, the redshifts are 
as accurate as $0.0005$. For four quasars that have been observed by 
SDSS, 
we use the average of three redshifts (pipeline redshifts, PCA redshifts 
and \ciii\ redshifts).
The systemic 
redshift of SDSS2346$-$0016 was derived in \cite{zheng15}, as
discrepancy was found and studied. The redshift of HS1024+1849 is estimated as described above. For the remaining five quasars at the bottom of Table \ref{tbl-data}, their 
redshifts and errors are from various literatures. When we derive quasar ages
from the proximity zones, uncertainties in both measurements and redshifts (see
the shaded regions in Figures \ref{fig-mosaic3}$-$\ref{fig-1024}) are included. 
When comparing the \he\ and  \h\ zone sizes, the redshift errors are not 
included since uncertainties in systemic redshift affect both proximity zones in the same manner. 

\subsection{External Effect}\label{sec_ext}

When a long line of sight toward a distant quasar intercepts the proximity zone of another
foreground quasar, a surge in flux may be observed in both the \he\ and \h\ spectra 
\citep{syphers14}. While it is common that a spectrum of \lya\ forest lines displays such voids,  
they may be confused as intrinsic if two quasars are within 20 Mpc in 
separation. The proximity profile of an external source is largely symmetrical, namely
a spectral void with wings toward both the longer and shorter wavelengths.
In optical spectra, if the strength of \lya\ forest lines declines and then recovers,
this is likely an external effect. We use this criterion to reject large zone sizes in 
several quasars. However, rejecting an external effect in a \he\ proximity profile is difficult 
because the \he\ counterparts of \lya\ forest lines are so strong that they make significant troughs
in the spectra of low resolution. The \he\ proximity profile reflects a gradual decline of flux 
towards larger radii. If a significant spectral void is found beyond the characteristic radius, 
it is considered as external, as illustrated in Panel {\it e} of Figure \ref{fig-mosaic3}.

\section{DISCUSSION}\label{sec_dis}

The time variations of IGM (intergalactic medium) reionization are governed by the two 
opposing processes of reionization and recombination \citep{meiksin}. For 
hydrogen:\newline \begin{eqnarray}
\frac{dx_{\rm HI}}{dt} &=& -x_{\rm HI}\Gamma_{\rm HI}
+x_{\rm HII}n_e\alpha_A, \nonumber\\
\frac{dx_{\rm HII}}{dt} &=& -\frac{dx_{\rm HI}}{dt},
\label{eq:phionizeH}
\end{eqnarray}

\noindent where \frach\ is the fractional population of neutral hydrogen, \frache\ that of 
ionized hydrogen, $n_e$ the electron density, $\alpha_A$ the case-A recombination coefficient, 
and $\Gamma_{\rm HI}$ the hydrogen-ionization rate per neutral hydrogen atom.

Similarly, for helium: 
\begin{eqnarray}
\frac{dx_{\rm HeI}}{dt}  &=& -x_{\rm HeI} \Gamma_{\rm HeI} + x_{\rm HeII} n_e\alpha_{\rm HeII}, \nonumber \\
\frac{dx_{\rm HeII}}{dt}&=& -\frac{dx_{\rm HeI}}{dt} - \frac{dx_{\rm HeIII}}{dt}, \nonumber \\
 \frac{dx_{\rm HeIII}}{dt}&=&  x_{\rm HeII} \Gamma_{\rm HeII} - x_{\rm HeIII} n_e\alpha_{\rm HeIII}, \label{eq:phionizeHe}
\end{eqnarray}

\noindent where $x_{\rm HeI}$ is the fractional population of neutral helium, \frache\ 
that of singly ionized 
helium, $x_{\rm HeIII}$ that of fully ionized helium,   
$\alpha_{\rm HeII}$ and $\alpha_{\rm HeIII}$ the total recombination
rates to all levels of HeI and \he, respectively, and 
$\Gamma_{\rm HeI}$ and $\Gamma_{\rm HeII}$ the respective photoionization rates. 
When the ionization term is suddenly increased, the equilibration scale toward a higher state is 
$t_{eq} = \Gamma^{-1}$. When this terms becomes insignificant, the time scale of reaching a new balance 
is $t_{rec}=(\alpha n_e)^{-1}$.

\subsection{Development of \he\ Ionization Zone}

The expansion of a quasar's \he\ proximity zone should be slower than its hydrogen counterpart
for two reasons: (1) a significant portion of intergalactic singly ionized helium exists at 
$z \gtsim 3$ and 
(2) the quasar's \he-ionizing flux at 4 Ry is considerably weaker than that at 1 Ry. 
As the ionizing photons from a quasar propagate though  intergalactic space, 
the enhanced ionization level would respond within an equilibration time scale. 
Since the size of a 
proximity zone $R$ is proportional to the square root of the quasar age.
At the edge of a proximity zone, this time scale of \h\ equilibration is 
$\sim 10^4$ yr \citep{meiksin,khrykin}.
While this time scale is too small to make a meaningful observable difference 
for \h, the \he\ equilibration time is considerably longer, on the order of Myr as 
$\Gamma_{\rm He II}$ is smaller by 2-3 orders of magnitude \citep{haardt12}.

\cite{khrykin} simulate
the \he\ ionization and find that the proximity profiles are different for various quasar ages,
in spite of a degeneracy between the \frache\ fraction and the ages. They suggest that the signature 
of quasar ages can be found up to about 30 Myr, or about 10 Mpc. 
This will serve as a gauge of the \h\ counterpart.

\subsection{\h\ Proximity Zone and Quasar Age}\label{sec-age}

Are the observed sizes of \h\ proximity zones relevant to quasar ages? 
To seek an answer, we compare the \h\ proximity zones with their \he\ counterparts. 
\he\ proximity zones are deemed as more sensitive and reliable probes than 
their \h\ counterparts, which can only be estimated indirectly.
Some simulations assume an infinite speed of light, because the transmission spectrum is along 
the line of sight. If this is the case, the effect of quasar radiation should be observed to the 
full \h\ zone size for a given quasar luminosity.

As shown in Figure \ref{fig-norm}, there are four data points of normalized 
proximity zones that are smaller than 0.2 for \he\ and 0.3 for \h. 
If the current models are correct, the ordinate values for these data points should be at 1 or higher, 
which are not observed. This provides clear evidence that the sizes of \h\ 
proximity zones for young quasars are small.
Only one data point in Figure \ref{fig-heh} shows high values of \he\ and \h\ proximity zones 
($>12$ Mpc), suggesting that it is likely an old quasar. Four other data points show large \he\
proximity zones ($> 12$ Mpc), but moderate \h\ zone sizes, suggesting that large spectral voids
at large distances from these quasars may not be attributed to a foreground quasar near the line of sight.
Overall, the results suggest that the sizes of \he\ and \h\ proximity zones are correlated, and they 
are indicative of quasar ages.

To estimate quasar ages from \he\ proximity profiles, we carry out simulations using the 
time-dependent Equations (1) and (2). The bin size is 0.03 Mpc, and the density fluctuation follows a 
lognormal distribution \citep{bi} in the range of $> 0.1 \times$ average IGM density.
At large radial distances, this nominal density is $1.85 \times 10^{-5}$ \cc\ for $z=2.885$.
For \qa, we assume a high luminosity of $\dot N_\gamma = 2\times 10^{58}$ \s\ above 1 Ry
and no time delay along a line of sight. 
With an assumed background radiation field of $\Gamma_{\rm HeII} = 10^{-14.3}$ 
\s, \he\ has been largely ionized. From a radial distance of 0.5 Mpc outward, the ionizing flux 
at each distance is calculated at energy bins above 1 Ryd. The time dependence of \he\ and \h\ populations is calculated over 
a range of quasar ages between $0.0001 - 100 $ Myr. 
The optical depths at different ionization energy levels are then 
updated before the next round at a slightly larger radial distance.

\begin{figure}
\plotone{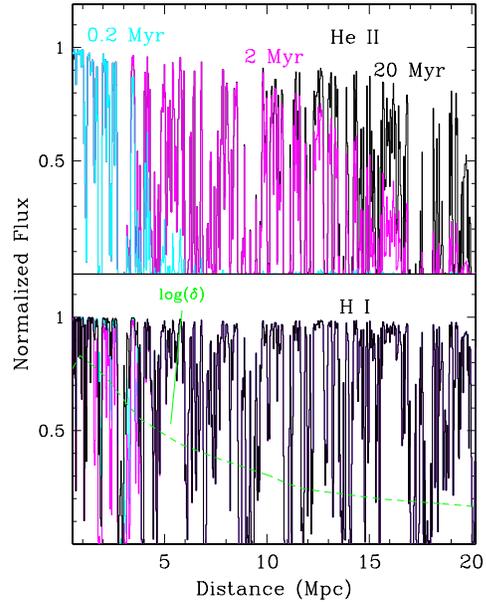}
\figcaption{Simulated \he\ proximity profiles at different ages of quasar \qa. 
The light-travel effect is not included when viewed along the line of sight. 
The IGM overdensity increases from $\delta \approx 1.5$ at 20 Mpc to 5 at 3 Mpc, as marked by a 
green logarithmic curve in the lower panel.
A high \he\ ionization rate of $2\times 10^{57}$ \s\ 
and a UV slope of $\alpha=-1.7$ are assumed. The UVB flux is $\Gamma_{\rm HI} = 10^{-12}$ and
$\Gamma_{\rm HeII} = 10^{-14.3}$ \s. The IGM temperature is $ 10^4$. 
A cyan curve represents a quasar age of 0.2 Myr, magenta 2 Myr,  
and black 20 Myr.
\label{fig-sim1}}
\end{figure}

Since quasars are believed to form in the massive galactic halos, an IGM overdensity in the quasar 
vicinity affects the sizes of both \he\ and \h\ proximity zones.
\cite{guimaraes} found a reduction of proximity effect for luminous quasars and suggested an overdensity 
factor of $2-5$ between radial distances of $\approx 7-2$ Mpc. 
We adapt the density profile from Figure 12 of that paper. As shown in Figure \ref{fig-sim1}, the ionization front 
expands fast under the high luminosity, at an approximate rate of $R \propto \sqrt{t}$.

\begin{figure}
\plotone{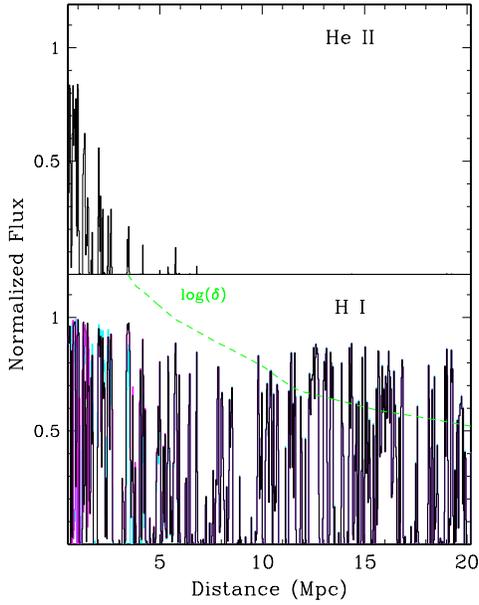}
\figcaption{Simulated \he\ proximity profiles at different ages of quasar \qa, with $5\times$ 
overdensity from $\delta \approx 3.5$ at 20 Mpc to 25 at 3 Mpc.
See caption for Figure \ref{fig-sim1}. The \he\ proximity zone becomes small, while deficiency in 
\h\ flux is significant.
\label{fig-sim2}}
\end{figure}

We carry out simulations 
at higher overdensities. As shown in Figure \ref{fig-sim2}, 
at an even higher overdensity ($5\times$ the density profile), the \he\ proximity zone 
indeed becomes small, but strong densities cause significant deficiency in the \h\ counterparts. Within 
the distance of 20 Mpc, no weak absorption lines are present. Therefore, very high overdensities seem
difficult to match both the \he\ and \h\ spectra.
 
\subsection{Episodic Quasar Bursts?}

\begin{figure}
\plotone{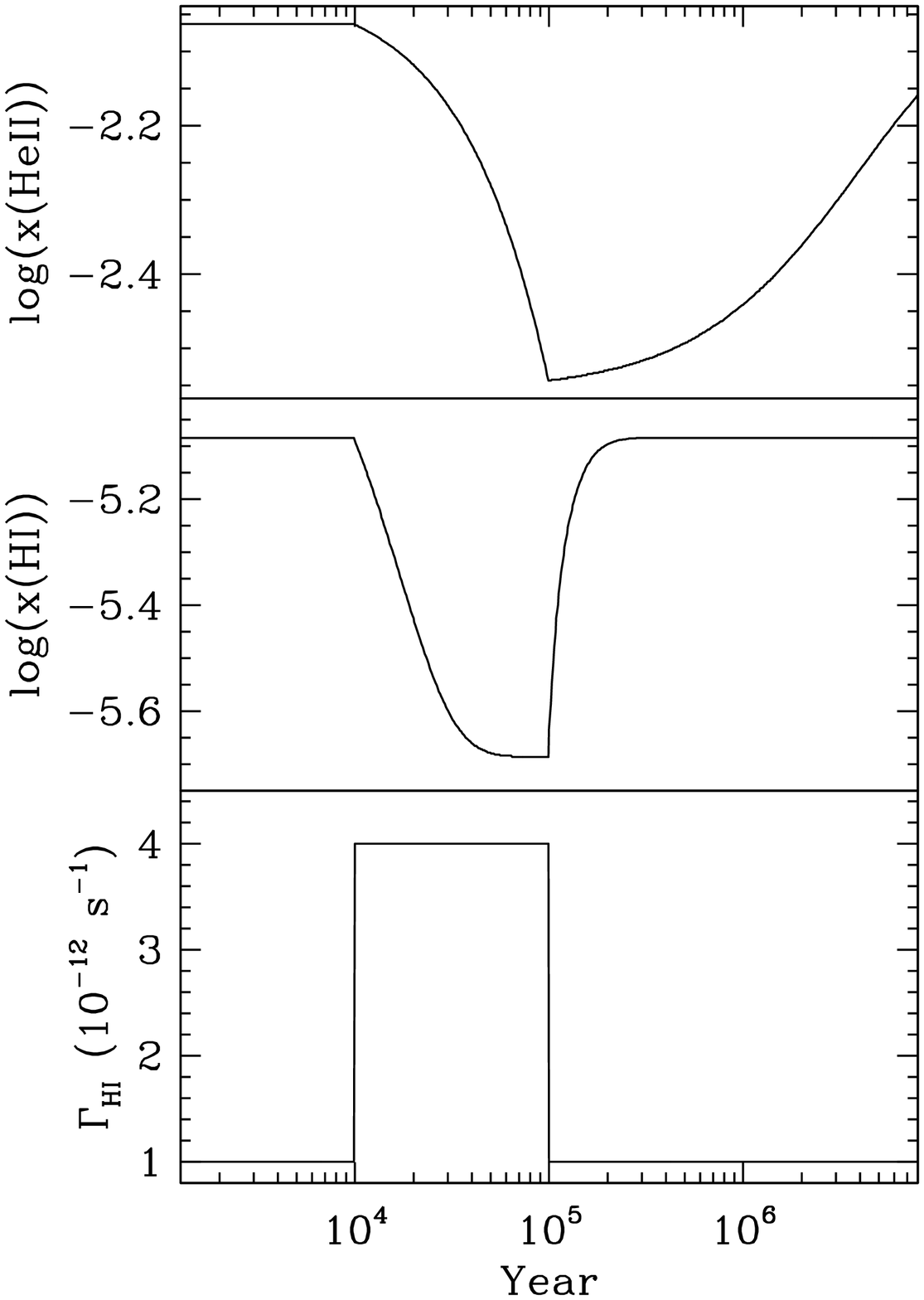}
\figcaption{Simulated response of \he\ and \h\ fractional populations to a quasar burst.
The UVB is assumed to be at $\Gamma_{\rm HI}=10^{-12}$ \s\ at 1 Ry and $\Gamma
_{\rm HeII}= 10^{-14.3}$ \s\ at 4 Ry.  
The distance to the quasar is
$\approx 0.57\ R_{\omega=1}$ so that the quasar flux is $\Gamma^{\rm Q}_{\rm HI} \approx 
3 \times 10^{-12}$ \s\ at 1 Ry and $\Gamma^{\rm Q}_{\rm HeII} \approx 3 \times 10^{-13}$ 
\s\ at 4 Ry. 
The fractional population of hydrogen responds within a time scale of 
$\sim 10^{4}$ yr, and that of \he\ considerably longer.
\label{fig-time}}
\end{figure}

A correlation between the \he\ and \h\ proximity zones implies that the expansion 
of \h\ ionizing front at $z\sim 3$ is largely sub-luminal. For the young quasars with small \he\ 
proximity zones, their \h\ counterparts are also small.  To our knowledge, this is 
not in agreement with current models.
One possible explanation is that quasar activities are flickering instead of steady. 
For the IGM at $z\sim 3$, the reionization term from a UVB field $\Gamma_{\rm HI} \sim 10^{-12}$ is normally balanced by 
the recombination term at \frach\ $\sim 10^{-5}$. Under the enhanced radiation from a quasar
and after a period of equilibration time, this balance is shifted to a lower level of \frach.
After the quasar radiation is switched off or enter a low state, the \frach\ value will 
increase back to its original level. While the characteristic time
scale of recombination is very long, it takes only a small 
fraction of that time to reach a new level of \frach$ \sim 10^{-5}$. At a \frach $<< 1$, 
Equations (2)  become $  {dx_{\rm HI}}/{dt}  \simeq n_e \alpha_A$, and the time for a rebalance is
$\Delta t \simeq \Delta x_{\rm HI} t_{rec}$, approximately $4\times 10^{4}$ yr for 
$\Delta x_{\rm HI} \sim 10^{-5}$. This trend is illustrated in Figure \ref{fig-time} for one single 
quasar burst. The response time for \he\ is considerably longer because \frache$ >> $\frach, 
therefore the effect for \he\ is accumulative over the entire quasar lifetime, and the \h\
effect responds to the latest quasar burst. 
Combining with the correction along lines of sight, it may be possible to see both the 
\he\ and \h\ proximity zones on a similar scale and are suggestive of quasar ages, namely 
distinguishing young and old ages.

Quasar flickering has been suggested by \cite{ciotti,novak}. 
\cite{schawinski} find evidence for a typical AGN phase of $10^5$ yr.
\cite{kirkman} study the transverse \h\ proximity effect in 130 quasar pairs and 
suggest episodic quasar lifetimes of $\sim 1 $ Myr.
Recently \cite{khrykin19} suggest that the small proximity zones in several quasars may be consistent 
with  time-dependent quasar activities. It is possible that, during the early stage of quasar 
activities,
flickering is common, and their \he\ and \h\ proximity zones are small. 
For old quasars, their activities are steadier, making large zones.
It is noted that this scenario is overly simplified, and 
more work in both observations and theoretical models is needed. 

\section{CONCLUSION}

We compare the \he\ and \h\ proximity zones in 15 quasars. 
While the sample is small, the high sensitivity of \he\ absorption and the high quality of 
optical data offer a unique chance to probe the propagation of \h\ proximity 
zones over quasar lifetimes. Our results suggest that
(1) for the four quasars with small \he\ proximity zones, no large \h\ proximity effect is
present.
(2) for the four quasars with large \he\ proximity zones ($R_{\rm HeII} \gtsim R_{\omega({\rm HeII})=1}$), their \h\ 
counterparts are also large ($R_{\rm HI} \gtsim 0.8 R_{\omega=1}$), suggesting that both \he\ and \h\ proximity zones 
are related to quasar ages.
(3) there is a significant correlation between \h\ and \h\ proximity-zone sizes, hence both may be indicative 
of quasar ages. \h\ proximity zones are often smaller than their \he\ counterparts. 
These properties remain significant even after normalizing for quasar luminosity
and are not in apparent agreement with the current models. The overdensity in the quasar vicinity 
reduces the proximity effect, but simulations with very high overdensities do not match both the observed
\he\ and \h\ spectra. A possible explanation is that quasar activities are episodic, resulting in 
periods of ionization and recombination
for both \he\ and \h\ in the vicinity. The \h\ ionization responds rapidly with the varying flux, but 
the \he\ ionization responds slowly over the entire lifespan of a quasar. 

Our model suggests that a \he\ proximity profile along the line of sight develops fast, 
reaching $\gtsim 10$ Mpc due to the first 1 Myr of quasar lifetime.
Quasars that display small \he\ and/or \h\ proximity zones of $\ltsim 3$ Mpc should be considerably 
younger than 1 Myr. It is emphasized that the proximity effect along lines of sight may 
provide only the upper limits to quasar ages, and the actual quasar ages may be considerably
smaller. Nevertheless, we should able to identify a group of young quasars from their small 
\h\ proximity zones, similar to that of \he\ counterparts.

\acknowledgments

This research is based on observations made with the NASA/ESA Hubble Space Telescope,
obtained from the Mikulski Archive for Space Telescopes (MAST) at the Space Telescope Science Institute, 
which is operated by the Association of Universities for Research in Astronomy, Inc., under NASA contract
NAS 5-26555. 
It has made use of the Keck Observatory Archive (KOA), which is operated by the W. M. Keck 
Observatory and the NASA Exoplanet Science Institute (NExScI), under contract with the National Aeronautics 
and Space Administration. It is also based on observations collected at the European Southern Observatory under ESO programmes 
65.O-0693(A), 68.A-0230(A), 083.A-0421(A) and 095.A-0654(A), obtained from the ESO Science Archive Facility, 
under request numbers 346827, 389941,  389967 and 389972-389977. 
We thank Robert Simcoe for allowing us to quote his unpublished data.

\clearpage

\begin{deluxetable}{llcccc}
\tablecaption{Quasar Redshifts and Observations\label{tbl-data}}
\tablewidth{0pt}
\tighttable
\tablehead{
\colhead{Name} &
\multicolumn{1}{c}{Redshift} &
\colhead{Optical} &
\colhead{Exp. (sec)} &
\colhead{UV/HST} &
\colhead{Exp. (sec)}
}
\startdata 
\qa & $2.885\pm 0.005$ \tablenotemark{a} & VLT/UVES & 28455 & COS/G130M & 69852 \\ 
\qb & $2.748 \pm 0.005$ \tablenotemark{b}& Keck/HIRES & 27632& COS/G130M & 18137 \\ \qc & $3.2860 \pm 0.0005$ \tablenotemark{c}& Keck/HIRES & 15595 &  COS/G130M & 21995 \\ 
SDSSJ0915+4756 & $3.341 \pm 0.005$ \tablenotemark{d}& Keck/HIRES & 10800 & COS/G130M & 26864\\ 
SDSSJ1237+0126 & $3.149 \pm 0.007$  \tablenotemark{d}& VLT/UVES & 47600 & COS/G140L & 3382  \\ 
SDSSJ0936+2927 & $2.925 \pm 0.001$ \tablenotemark{d}& Keck/HIRES & 14400 & COS/G140L & 2943\\  
SDSSJ1253+6817 & $3.475 \pm 0.002$ \tablenotemark{d}& Keck/HIRES & 15000 & COS/G140L & 14096\\  
HS1024+1849 & $2.855 \pm 0.008$ \tablenotemark{e}& VLT/UVES & 3000 & COS/G130M & 28689\\
SDSS2346$-$0016 & $3.511\pm 0.003$ \tablenotemark{f}& Keck/HIRES & 18900  & COS/G140L & 4679 \\ SDSSJ1711+6052 & $3.835 \pm 0.011$ \tablenotemark{g}& Keck/HIRES & 25200  & COS/G140L & 23951 \\  
QSOJ1630+043 & $3.81 \pm 0.006$ \tablenotemark{g}& VLT/UVES & 12000 & COS/G130M & 40919\\
\qd & $3.70 \pm 0.01$ \tablenotemark{g} & VLT/UVES & 29865 & COS/G130M & 44882 \\  
PC0058+0215 & $2.89$ \tablenotemark{h}& VLT/UVES & 27000  & COS/G140L & 6212\\
PKS1935$-$692 & $3.185  $ \tablenotemark{i}& VLT/UVES & 18112 & GHRS/G140L & 82579\\
QSOJ2157+2330 & $3.143$ \tablenotemark{j}& VLT/UVES & 15000 & COS/G140L & 5524\\ 
\hline \enddata
\tablenotetext{a}{\cite{reimers}, $^{\rm b}$\cite{syphers13}, $^{\rm c}$\cite{syphers14},
$^{\rm d}$SDSS DR12 quasar catalog \citep{dr12}, $^{\rm e}$this work, 
$^{\rm f}$\cite{zheng15}, $^{\rm g}$\cite{khrykin19}, $^{\rm h}$\cite{worseck16}, 
$^{\rm i}$\cite{anderson}, $^{\rm j}$HST/GO proposal 13013, \cite{hst13013}.}
\end{deluxetable}
 
\begin{deluxetable}{lcccc}
\tablecaption{Measurements and Estimates of Proximity Zones\tablenotemark{a}\label{tbl-zone}}
\tablewidth{0pt}
\tighttable
\tablehead{
\colhead{Name} &
\colhead{$R_{\rm HeII}$} &
\colhead{$R_{\omega({\rm HeII})=1}$\tablenotemark{b}} &
\colhead{$R_{\rm HI}$} &
\colhead{$R_{\omega=1}$\tablenotemark{c}}
}
\startdata 
\qa & $0^{+1.3}_{-0}$ & 30.1 & $0.8 \pm 0.8$& $19.8 $ \\
\qb &  $5.8 \pm 1.0$ & 32.4 &$6.1 \pm 1.3$& $21.3$ \\
\qc & $ 12.7^{+0.7}_{-2.0} $ & 18.3 & $7.1 \pm 1.4$& $12.0 $ \\
SDSSJ0915+4756 &$2.1\pm 0.4$ &15.8 & $5.3\pm 1.0 $& $10.4$ \\ SDSSJ1237+0126 & $7.4 \pm 1.1 $ & 9.4 &$8.5 \pm 1.9$& $6.1$ \\ SDSSJ1253+6817 & $10.3 \pm 1.2 $ & 12.0 & $6.1 \pm 0.7$ & $7.9$ \\ HS1024+1849 & $12.3\pm 3.5 $ & 14.2 & $16.0 \pm 2.5$ & 9.3  \\
SDSSJ0936+2927 & $16.7 \pm 1.7$ & 12.2 &$6.9 \pm 1.2$& $8.0$ \\
SDSS2346$-$0016 & $2.8\pm 0.3 $ & 17.0 & $2.6 \pm 0.7$ & 11.1 \\
SDSSJ1711+6052 & $2.8 \pm 0.5$ & 8.9 &$1.2 \pm 1.2$& $ 5.9 $ \\
QSOJ1630+043 & $7.6\pm 0.7 $ & 19.8 & $4.8 \pm 0.6$ & 13.0 \\
\qd & $2.1 \pm 0.6 $ & 15.4 &$2.2 \pm 0.9$& $10.1$ \\
PC0058+0215 & $8.0\pm 1.3$ & 8.3 & $4.9\pm 0.8$ & 5.4   \\
PKS1935$-$692 & $15.6 \pm 1.4$ & 16.6  &$6.3 \pm 0.7$ & 10.9   \\
QSOJ2157+2330 & $18.1 \pm 1.5$ & 14.3  &$7.9 \pm 1.9$ & 9.4   \\
\hline
\enddata
\tablenotetext{a}{In units of Mpc, excluding the effect of redshift uncertainties.}
\tablenotetext{b}{Characteristic radius at which the quasar ionizing flux above 
4 Ry is equal to that of the UVB of $\Gamma_{\rm HeII}=10^{-14.3}$ \s.}
\tablenotetext{c}{Characteristic radius at which the quasar ionizing flux above 
1 Ry is equal to that of the UVB of $\Gamma_{\rm HI}=10^{-12}$ \s.} 
\end{deluxetable} 

\end{document}